**Strong Dzyaloshinskii-Moriya Interaction and Origin of Ferroelectricity in $Cu_2OSeO_3$**

J. H. Yang[1], Z. L. Li[1], X. Z. Lu[1], M.-H. Whangbo[2], Su-Huai Wei[3], X. G. Gong[1*],

and H. J. Xiang[1*]

[1] Key Laboratory of Computational Physical Sciences (Ministry of Education), State Key

Laboratory of Surface Physics, and Department of Physics, Fudan University, Shanghai 200433,

P. R. China

[2] Department of Chemistry, North Carolina State University, Raleigh, North Carolina 27695-8204,

USA

[3] National Renewable Energy Laboratory, Golden, Colorado 80401, USA

**Abstract**

By performing density functional calculations, we investigate the origin of the skyrmion state and ferroelectricity in $Cu_2OSeO_3$. We find that the Dzyaloshinskii-Moriya interactions between the two different kinds of Cu ions are extremely strong and induce the helical ground state and the skyrmion state in the absence and presence of magnetic field, respectively. On the basis of the general model for the spin-order induced polarization, we propose that the ferroelectric polarization of $Cu_2OSeO_3$ in the collinear ferrimagnetic state arises from an unusual mechanism, i.e., the single-spin-site contribution due to the spin-orbit coupling.

PACS numbers: **75.85.+t, 71.20.-b, 75.30.Et, 75.50.Gg**

Skyrmions are topologically protected field configurations with particle-like properties, and were predicted to exist in magnets [1]. Recently neutron scattering and Lorentz transmission electron microscopy measurements showed that skyrmions with diameters of about 20~90 nm are induced from a helical magnetic ground state by an external magnetic field in bulk magnets MnSi [2] and FeCoSi [3] with chiral crystal structure. Electron flow with low current density can make skyrmions move, suggesting that skyrmion crystals may have potential applications in high density magnetic storage devices. Recently, Seki $et$ $al$. discovered magnetoelectric skyrmions in an insulating chiral-lattice magnet $Cu_2OSeO_3$ [4]. This phenomenon suggests the possibility of manipulating skyrmions by electric field without destroying them.

Bulk $Cu_2OSeO_3$ has the same space group (cubic and chiral) $P2_13$ as do the B20 MnSi alloys. $Cu_2OSeO_3$ has two kinds of $Cu^{2+}$ ions, namely, Cu-I and Cu-II with the [Cu-I]:[Cu-II] ratio of 1:3. The Cu-I atoms form $CuO_5$ trigonal bipyramids, and the Cu-II atoms $CuO_5$ square pyramids [see Fig. 1(a)]. A previous magnetic study of $Cu_2OSeO_3$ [5] with magnetic field of 20 kOe indicated a collinear ferrimagnetic order at 58.8 K between the Cu-I and Cu-II spins. The [77]Se nuclear-magnetic-resonance (NMR) study [6] with a single crystal $Cu_2OSeO_3$ also suggested a transition from the high-temperature paramagnetic to a low-temperature ferrimagnetic phase under a magnetic field of 14 T. Recently, Seki $et$ $al$. found that the spin ground state of bulk $Cu_2OSeO_3$ is a long wavelength helical state, and an external magnetic field of about 1 kOe can induce a skyrmion state in bulk $Cu_2OSeO_3$, and more easily in thin-film $Cu_2OSeO_3$ in a certain temperature range. Subsequently, neutron scattering and magnetization measurements confirmed the helimagnetic spin ground state and the skyrmion lattice phase induced by an external magnetic field [7]. By analogy with B20 MnSi, it was proposed [4,7] that Dzyaloshinskii-Moriya (DM) interaction is responsible for the occurrence of the helimagnetic spin ground state and the skyrmion lattice phase under magnetic field. So far, the detailed microscopic origin of these interesting magnetic properties in $Cu_2OSeO_3$ is unclear.

The magnetoelectric and multiferroic properties in $Cu_2OSeO_3$ are interesting [8]. Bos *et al*. [8] found that significant magneto-capacitance develops in the ordered state below the transition temperature, and that $Cu_2OSeO_3$ shows no measurable structural distortion down to 10 K. These suggest that the magnetoelectric coupling may not proceed via a spontaneous lattice distortion below the magnetic ordering temperature. The electric polarization measurements by Seki *et al*. reveal that the ferrimagnetic, the helimagnetic, and even the skyrmion lattice spin state can magnetically induce nonzero electric polarization. In particular, the ferroelectric polarization P saturates (around 17 μC/m$^2$) in the ferrimagnetic state when the magnetic field is applied along [111], and P is parallel to H when the latter is along [111] in these magnetic phases. The ferroelectric polarization in the collinear ferrimagnetic state is apparently puzzling because neither the symmetric exchange striction mechanism [9,10] nor the general spin current model [11,12] can provide a non-zero electric polarization.

In this Letter, we explore the magnetic structure and ferroelectric properties of $Cu_2OSeO_3$ on the basis of density functional theory (DFT) and spin-order induced ferroelectric polarization model. We find that the DM interactions between the two different kinds of Cu ions are very strong and give rise to the helical ground state and the skyrmion state under an external magnetic field, and that the ferroelectric polarization of $Cu_2OSeO_3$ in the ferrimagnetic state originates from an unusual mechanism, i.e., the single-spin-site contribution brought about by spin-orbit coupling (SOC).

The density of states (DOS) obtained for the ferrimagnetic state from the DFT+U calculations [see the Supporting Information (SI) for details], presented in Fig. 2(a), show that the system is insulating with a band gap about 2.1 eV. There is a hole in the $d_{z2}$ orbital for the Cu-I ion [see Fig. 2(b)] but in the $d_{x2-y2}$ orbital for the Cu-II ion (not shown) so that the magnetic orbitals of the Cu-I and Cu-II ions are described by the $d_{z2}$ and $d_{x2-y2}$ orbitals, respectively.

To investigate the magnetic properties of $Cu_2OSeO_3$, we evaluate the symmetric spin exchange interactions between the $Cu^{2+}$ ions using the four-state energy-mapping analysis [13]. We consider all superexchange interactions [See Fig. 1(b)] and all super-superexchange interactions with O...O distance less than 2.7 Å. The superexchange interaction between Cu-I and Cu-II are antiferromagnetic ($J_2 = 6.534$ meV and $J_4 = 0.900$ meV. Here the spin exchange refers to the effective spin exchange "$J_{ij}S_iS_j$" for any pair of spin sites i and j.), while the superexchange interaction between Cu-II ions ferromagnetic ($J_1 = -1.132$ meV and $J_3 = -3.693$ meV). Among the super-superexchange interactions, only one is important, i.e., the exchange $J_5$ between Cu-I and Cu-II ions with a distance of 6.35 Å, which is antiferromagnetic (0.984 meV). The antiferromagnetic interactions between Cu-I and Cu-II ions and the ferromagnetic interactions between Cu-II ions explain the observed ferrimagnetic (or nearly ferrimagnetic) spin order in $Cu_2OSeO_3$. The sign and magnitude of the exchange interactions can be understood within the superexchange theory [14] by evaluating the hopping parameters using the maximally localized Wannier functions [15,16] (see SI). Belesi *et al.* proposed that $J_2 = J_4 = 68$ K, and $J_1 = J_3 = -50$ K [6] by fitting the measured magnetization data with the mean-field theory assuming that all the superexchange interactions between the Cu-I and Cu-II ions are identical, and so are those between the Cu-II ions. Our calculations show that the antiferromagnetic exchanges between Cu-I and Cu-II are stronger than the ferromagnetic exchanges between the Cu-II ions, but the $J_2/J_4$ and $J_3/J_1$ ratios are much larger than 1.

To verify the experimental indication that the DM interactions between Cu ions might play an important role on the magnetic properties of $Cu_2OSeO_3$, we compute the DM interactions for the four superexchange paths ($J_1$-$J_4$) by using the energy-mapping analysis [13] to find that the DM interactions between the Cu-I and Cu-II ions are strong, namely, $\mathbf{D}_2$ = (-1.120, 1.376, -0.300) meV and $\mathbf{D}_4$ = (0.490, -1.238, -1.144) meV. $\mathbf{D}_4$ is much larger than $J_4$ in magnitude with $|D_4/J_4|$ = 1.95. It should be noted that $Cu_2OSeO_3$ has the largest $|D/J|$ value to the best of our knowledge, because usually $|D/J|$ is expected to be less than 0.05 [17]. As shown in Fig. 1(b), the DM vectors are almost perpendicular to the bond. The DM interactions between Cu-II ions are somewhat weaker with $|D_1/J_1|$ = 0.39 and $|D_3/J_3|$ = 0.14. As expected from the symmetry argument, the DM interactions between nonequivalent Cu ions are larger than those between equivalent Cu ions. The microscopic mechanism for the large DM interactions between Cu-I and Cu-II ions is similar to what was pointed out in the study of $CaMn_7O_{12}$ [10]: The SOC leads to the mixing of the Cu-I $d_{z2}$ orbital with the $\{d_{xz},d_{yz}\}$ orbitals, then the resulting hole in the $\{d_{xz},d_{yz}\}$ orbitals can hop to the Cu-II $d_{x2-y2}$ orbital because the Cu-O-Cu angle is close to 90°, and finally the hole hops back to the Cu-I $d_{z2}$ orbital because $t_2$ and $t_4$ are large. A similar three-hopping process also occurs for Cu-II. Our first principles calculations show that both processes are important for the DM vectors.

Experimentally, the ground state is a long-wavelength helical state. To see the effect of the DM interactions on spin order, we consider the energy of the proper-screw spin spiral characterized by the magnetic propagation vector $\boldsymbol{q}$. In this case, the spin direction of the Cu ions can be described by $S_{k,\boldsymbol{R}} = \cos[2\pi\boldsymbol{q}\cdot(\boldsymbol{R}+\boldsymbol{r_k})+\varphi_k^0]\,\boldsymbol{e}_x^q + \sin[2\pi\boldsymbol{q}\cdot(\boldsymbol{R}+\boldsymbol{r_k})+\varphi_k^0]\,\boldsymbol{e}_y^q$, where $k$ refers to the 16 Cu atoms in the unit cell, $\boldsymbol{R}$ is the lattice vector, $\boldsymbol{r_k}$ is the position of site $k$ within the unit cell, $\varphi_k^0$ is the phase of the spin at site $k$ in the collinear ferrimagnetic state (i.e., $\pi$ for Cu-I and 0 for Cu-II), $\boldsymbol{e}_x^q$ and $\boldsymbol{e}_y^q$ are the two orthonormal vectors that are perpendicular to $\boldsymbol{q}$. The total spin interaction energy can be written as $H = \sum_{<ij>} J_{ij} \, \boldsymbol{S}_i \cdot \boldsymbol{S}_j + \boldsymbol{D}_{ij} \cdot (\boldsymbol{S}_i \times \boldsymbol{S}_j)$, where <ij> refer to the Cu pairs corresponding to $J_1$-$J_5$. Because the helical spiral order deviates only slightly from the ferrimagnetic order and thus $q = |\boldsymbol{q}|$ is very small, we compute the total energy difference between the helical state and the ferrimagnetic state up to the second order of $q$ to find $\delta E = 0.1584 q^2 - 0.0033 q$ (eV/unit cell). When $q = 0.0104$, the helical spiral state has the lowest energy, which is only 0.017 meV/unit cell lower than the ferrimagnetic state. The small energy difference explains why a small magnetic field can polarize the helical state to the ferrimagnetic state. Our results show that the DM interactions between Cu-I and Cu-II are mainly responsible for the helical spin ground state. Experimentally, the $q$ value is about 0.014 [7] or 0.018 [4], which is close to our predicted value. Interestingly, the total energy of the helical state depends only on $q$, but not the direction of $\boldsymbol{q}$ due to the cubic symmetry of the system. Experimentally, different $\boldsymbol{q}$ directions were observed: Seki *et al.* observed that $\boldsymbol{q}$ in thin film $Cu_2OSeO_3$ is along <110> direction, while the neutron scattering experiment [7] on bulk $Cu_2OSeO_3$ suggested that $\boldsymbol{q}$ is along <100>. Our results show that all the helical spin states with the same $q$ are degenerate (at least within the second order approximation), thus explaining the discrepancy between different experimental results. The high degeneracy of the helical spin ground state may also explain the occurrence of the skyrmion state under the magnetic field: At certain temperature, the skyrmion state may be stabilized by the entropy effect due to an "order-by disorder" mechanism.

Now we turn to the ferroelectric properties of $Cu_2OSeO_3$. In general, the spin order induced ferroelectric polarization can be written as $\boldsymbol{P}_t = \sum_i \boldsymbol{P}_s(\boldsymbol{S}_i) + \sum_{<ij>} \boldsymbol{P}_p(\boldsymbol{S}_i, \boldsymbol{S}_j) + \cdots$. This is similar to the cluster expansion approach [18] in the alloy theory. As discussed previously [10,12], the pair terms $\boldsymbol{P}_p$ include the symmetric exchange striction and SOC related spin current-like ($\boldsymbol{S}_i \times \boldsymbol{S}_j$) terms. For the collinear ferrimagnetic state of $Cu_2OSeO_3$, the sum of the symmetric exchange striction vanishes due to symmetry, as confirmed by the direct DFT calculation without including SOC. The ($\boldsymbol{S}_i \times \boldsymbol{S}_j$) terms also vanish due to the collinear nature of the magnetic order. Because of the time-reversal symmetry, the single site term can be written as:

$$\boldsymbol{P}_s(\boldsymbol{S}) = \boldsymbol{P}_{xx}S_x^2 + \boldsymbol{P}_{yy}S_y^2 + \boldsymbol{P}_{zz}S_z^2 + 2\boldsymbol{P}_{xy}S_xS_y + 2\boldsymbol{P}_{xz}S_xS_z + 2\boldsymbol{P}_{yz}S_yS_z$$

$$= (S_x, S_y, S_z)\begin{pmatrix} \boldsymbol{P}_{xx} & \boldsymbol{P}_{xy} & \boldsymbol{P}_{xz} \\ \boldsymbol{P}_{yx} & \boldsymbol{P}_{yy} & \boldsymbol{P}_{yz} \\ \boldsymbol{P}_{zx} & \boldsymbol{P}_{zy} & \boldsymbol{P}_{zz} \end{pmatrix}\begin{pmatrix} S_x \\ S_y \\ S_z \end{pmatrix} = S^t P_M S.$$

If the magnetic ion is at the center of the spatial inversion symmetry, the single site term is zero. In $Cu_2OSeO_3$, there is no spatial inversion symmetry, so the single site term can be nonzero. It should be noted that the single-site term is a consequence of SOC because the Hamiltonian without including SOC is invariant with the rotation of the spin.

Using the mapping method (see SI), we extract the coefficients of the single-site terms for Cu-I and Cu-II. As discussed in SI, we can determine $\boldsymbol{P}_{yy} - \boldsymbol{P}_{xx}$ and $\boldsymbol{P}_{zz} - \boldsymbol{P}_{xx}$, but not individual $\boldsymbol{P}_{xx}$, $\boldsymbol{P}_{yy}$, $\boldsymbol{P}_{zz}$. However, by enforcing the condition $\boldsymbol{P}_{xx} + \boldsymbol{P}_{yy} + \boldsymbol{P}_{zz} = 0$, which removes the spin independent contribution because $\frac{1}{4\pi}\int \boldsymbol{P}_s(\boldsymbol{S})d\Omega = \frac{1}{3}(\boldsymbol{P}_{xx} + \boldsymbol{P}_{yy} + \boldsymbol{P}_{zz})$, we find that the coefficients for the Cu-I ion is much larger than those for Cu-II. Our calculations show that the coefficient matrix for Cu-I in the local axis system [XYZ, see Fig. 2(d)] can be written as:

$$P_M = \begin{bmatrix} (126,38,148) & (38,-126,0) & (3,7,0) \\ (38,-126,0) & (-126,-38,148) & (-8,3,0) \\ (3,7,0) & (-8,3,0) & (0,0,-296) \end{bmatrix} \times 10^{-6} \text{ eÅ}.$$

Note that the matrix $P_M$ satisfies the local symmetry (C$_3$) of the Cu-I ion, and that $\mathbf{P}_{ZZ}$ is along the -Z direction, while $\mathbf{P}_{XX}$ is in the XZ plane. When the spin of Cu-I is along Z, the unoccupied spin minority d$_{z2}$ orbital (here the d orbital refers to the hybridized orbital which contains the ligand tails) will mix with the occupied spin majority d$_{xz}$ and d$_{yz}$ orbitals according to the SOC operator [19]. When the spin of Cu-I is along X, the unoccupied spin minority d$_{z2}$ orbital will mix with the occupied spin majority d$_{xz}$ and the occupied spin minority d$_{yz}$ orbitals. The PDOS plot of Fig. 2(b) shows that the spin minority d$_{yz}$ orbital has almost the same average energy as the spin majority d$_{yz}$ orbital. The spin orientation, which leads to a larger mixing of the unoccupied d$_{z2}$ orbital to the occupied manifold, depends on the detailed hybridization between the ligands and Cu ions. From the difference in the electron density between the ∥Z and ∥X spin cases, we find that the ∥Z spin case leads to more electrons occupying the d$_{z2}$ orbital, in agreement with the direction of $\mathbf{P}_{ZZ}$. We note that the d$_{yz}$ orbital is not symmetric with respect to the YZ-plane due to the 2p tails of the in-plane oxygens. Thus, $\mathbf{P}_{XX}$ is almost in the XZ plane. The microscopic mechanism for the electric polarization is similar to the spin-dependent hybridization mechanism proposed by Arima *et al.* [20，21].

With the coefficient matrix, we can obtain the total polarization of the system: $\boldsymbol{P}_t = \sum_i \boldsymbol{P}_s(\boldsymbol{S}_i)$. Let us assume that the system is always in the ferrimagnetic state but the spin axis $\boldsymbol{h} = (h_x, h_y, h_z) = (\sin\theta\cos\varphi, \sin\theta\sin\varphi, \cos\theta)$ can be arbitrary. By straightforward algebraic manipulation, we obtain for the total electric polarization: $\boldsymbol{P}_t = A(h_y h_z \boldsymbol{e}_x + h_x h_z \boldsymbol{e}_y + h_x h_y \boldsymbol{e}_z)$ with $A = -40.00 \ \mu\text{C/m}^2$. If the spin direction is confined within the *ab*-plane, i.e., $\theta = 90°$, then $\boldsymbol{P}_t = A\boldsymbol{e}_z\cos\varphi\sin\varphi$. We gradually rotate the spin axis from the *x* axis to -*x* axis within the *ab*-plane, and then compute the polarization using the DFT+U+SOC method. We find that the polarization is along the *c* axis, as expected. As shown in Fig. 3, the z component of the polarization from the direct first-principles calculations is consistent with the $-\sin 2\varphi$ dependence from the polarization model. Only when the spin axis is along <100>, the spin order induced polarization is zero. To see which ferrimagnetic state has the largest polarization, we maximize $|\boldsymbol{P_t}| = |A|\sqrt{\sin^4\theta\cos^2\varphi\sin^4\varphi + \cos^2\theta\sin^2\theta}$. When the spin axis is along <111>, the ferrimagnetic state has the largest polarization $\frac{\sqrt{3}}{3}|A| = 23.1$ µC/m$^2$, which is also confirmed by the DFT calculations: We gradually vary the spin axis in the [110] plane and find that the largest polarization occurs for [111] and [11-1] direction. When the spin axis is along [111], the computed polarization is 28.6 µC/m$^2$ (along the –[111] direction), which is close to the experimental result (about 17 µC/m$^2$) [4]. When the spin axis is reversed, i.e., along -[111] direction, both the direction and magnitude of the polarization do not change, in agreement with the experimental finding (see Fig. 3 of Ref. [4]). It should be noted that the change of the magnetization direction may change the direction of the polarization, but it is not possible to switch the electric polarization direction from –[111] direction to the [111] direction.

Experimentally [4], it was found that the polarization is proportional to M$^2$, where M is the magnetization, when the magnetic field is applied to switch the spin order from the helical state to the collinear ferrimagnetic order. The M$^2$ dependence is in agreement with our analysis because, if we assume that the spin components perpendicular to the field are random, the total polarization depends exactly on M$^2$ as a result of the second order nature of the single site term. Thus, the ferroelectricity in Cu$_2$OSeO$_3$ is due to the single-site term, similar to the case of Ba$_2$CoGe$_2$O$_7$ [22,23].

Our discussions presented above neglect the ion-displacement effect. Our test calculations show that the ion displacements only make the polarization larger (see SI), which differs from the TbMnO$_3$ case [24].

Work at Fudan was partially supported by NSFC, the Special Funds for Major State Basic Research, Pujiang plan, FANEDD, Research Program of Shanghai municipality and MOE. We thank Dr. Shinichiro Seki for the enlightening discussions.

e-mail: xggong@fudan.edu.cn

e-mail: hxiang@fudan.edu.cn

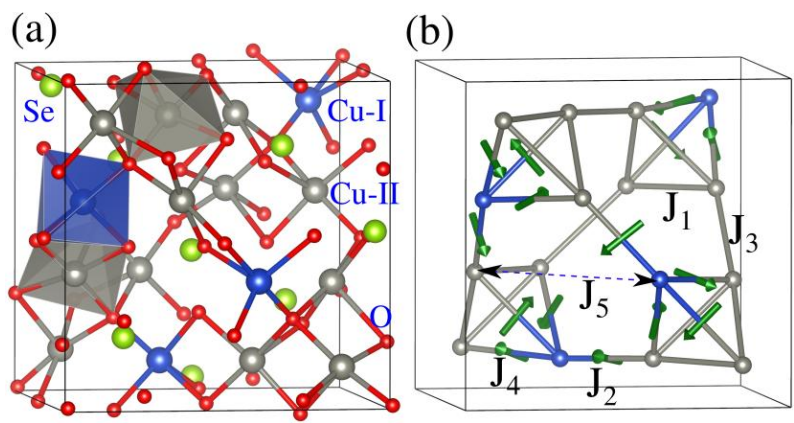

FIG. 1 (color online). (a) Crystal structure of $Cu_2OSeO_3$. One $\{Cu\text{-}I\}O_5$ trigonal bipyramid and two $\{Cu\text{-}II\}O_5$ square pyramids are shown using polyhedrons. (b) Exchange paths in $Cu_2OSeO_3$. The arrows illustrate the DM vectors ($\mathbf{D}_2$ and $\mathbf{D}_4$) between Cu-I and Cu-II ions. For clarity, O and Se ions are omitted.

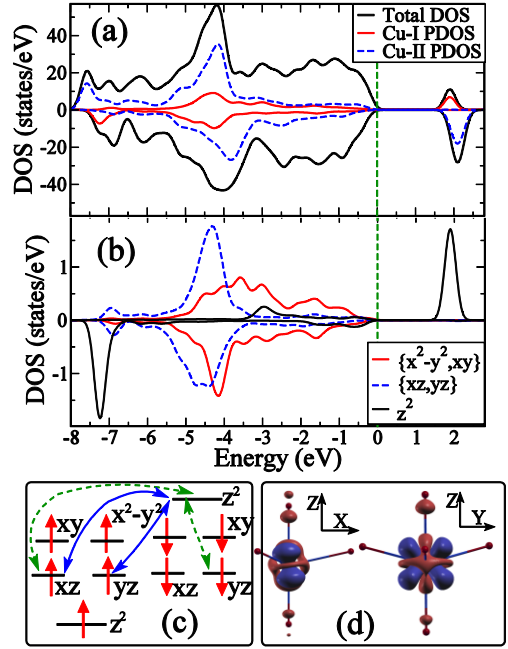

FIG. 2 (color online). (a) The density of states plot from the GGA+U calculations on the ferrimagnetic state. (b) The partial density of states for the Cu-I ion. The local coordination system XYZ shown in (d) is adopted. (c) The mixing between the unoccupied $d_{z^2}$ orbitals and occupied orbitals. The solid lines with two-headed arrows show the case of spin∥Z, and the dashed lines for spin∥X. (d) The electron density difference between the case of spin∥Z and the case of spin∥X.

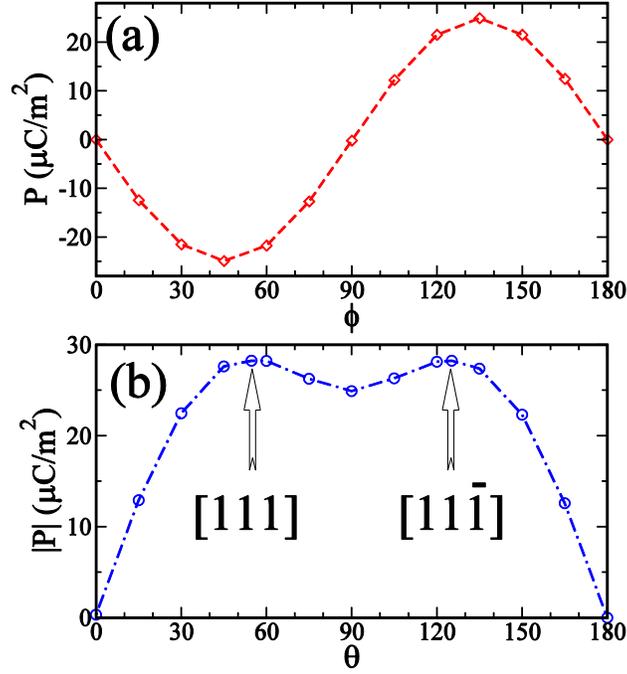

FIG. 3 (color online). (a) The z component of the electric polarization as a function of the angle (φ) between the spin axis and the x axis when the spin axis is rotated in the *ab*-plane. The polarization depends on φ as −sin(2φ). The ferrimagnetic state is adopted in the DFT+U+SOC calculation. The other two components of the electric polarization are zero in this case. (b) The magnitude of the electric polarization as a function of the angle (θ) between the spin axis and the z axis when the spin axis is rotated in the (110) plane. The system has the maximum polarization when the spin axis is along <111>.